\documentclass{elsart}

\usepackage{natbib}

 \usepackage{graphicx}

\usepackage{amssymb}

\begin{document}

\begin{frontmatter}


 \title{3D Studies of Neutral and Ionised Gas and Stars in Seyfert and Inactive Galaxies}
 \author{C.G. Mundell\thanksref{1}}
 \author{G. Dumas\thanksref{1}\thanksref{2}}
 \author{E. Schinnerer\thanksref{3}}
 \author{N. Nagar\thanksref{4}}
 \author{S. Haan\thanksref{3}}
 \author{E. Wilcots\thanksref{5}}
 \author{A.S. Wilson\thanksref{6}}
 \author{E. Emsellem\thanksref{2}}
 \author{P. Ferruit\thanksref{2}}
\author{R.F. Peletier\thanksref{7}}
\author{P.T. De Zeeuw\thanksref{8}}


 \address[1]{Astrophysics Research Institute, Liverpool J.M. Uni., Birkenhead, CH41 1LD}
\address[2]{CRAL-Observatoire, 9 avenue Charles Andr\'e, 69231 Saint Genis Laval, France}
\address[3]{Max-Planck-Institut f\"ur Astronomie, K\"onigstuhl 17, 69117 Heidelberg, Germany}
\address[4]{Uni. de Concepcion, Grupo de Astronomia, Casilla 160-C, Concepcion, Chile}
\address[5]{Dept. of Astronomy, University of Wisconsin, Madison, WI 53706, USA}
\address[6]{Dept. of Astronomy, University of Maryland, College Park, MD 20742, USA}
\address[7]{Kapteyn Astronomical Institute, Uni. of Groningen, Postbus 800, 9700 AV Groningen, the Netherlands}
\address[8]{Leiden Observatory, Postbus 9513, 2300 RA Leiden, the Netherlands}

\begin{abstract} We are conducting the first systematic 3D
spectroscopic imaging survey to quantify the properties of the atomic
gas (H{\sc i}) in a distance-limited sample of 28 Seyfert galaxies and a
sample of 28 inactive control galaxies with well-matched optical
properties (the {\bf VHIKINGS} survey). This study aims to address the
role of the host galaxy in nuclear activity and confront outstanding
controversies in optical/IR imaging surveys. Early results show
possible relationships between Seyfert activity and HI extent, content
and the prevalence of small, nearby gas-rich dwarf galaxies (M$_{\rm
HI}$ $\sim$10$^7$ M$_{\odot}$); results will be tested via rigorous
comparison with control galaxies. Initial results from our optical
followup study of 15 of our galaxies using the SAURON integral field
unit on the WHT suggest a possible difference between Seyfert and
inactive stellar and gaseous kinematics that support the conclusion
that internal kinematics of galaxies are the key to nuclear activity.

\end{abstract}


\end{frontmatter}

\section{Introduction}

Traditionally, the broader study of the formation, structure and
evolution of galaxies largely excluded Active Galactic Nuclei
(AGN: distant quasars and nearby Seyfert galaxies), whose energy
output is primarily powered by the release of gravitational potential
energy as material from the host galaxy is accreted by a central
supermassive black hole.  Recently, however, it has been recognised
that the growth and influence of the black hole, the
formation/evolution of galaxies, star formation and nuclear activity
at different cosmic epochs are intimately related
 \cite[e.g.][]{sr98,hop06,rob06}, and that most bulge-dominated galaxies
 today harbor relic black holes from an early quasar phase
 \cite[e.g.][]{geb00,mf01}. 

Local galaxies are well-established, so re-ignition of dormant black
holes is required to explain why ongoing activity is observed in only
10$-$20\% of galaxies, despite the ubiquity of black holes.  A key
question is whether the ignition mechanism is related to the galactic
host properties; numerical models predict large gas inflows due to
gravitational torques in non-axisymmetric potentials related to bars
or galaxy interactions \cite{bh96,bcs05,at05}, but the majority of optical/IR
{\em imaging} studies have failed to find a clear distinction between
Seyfert host and inactive galaxy morphology on a range of scales that
encompasses nearby companions and interactions, galactic bars and
nuclear spirals \cite[e.g.][]{mr97,ksp00,mart03}. However, recent
NICMOS imaging of the circumnuclear regions of 250 galaxies
\cite{hm04} found significant isophotal twists and disturbances
suggesting the possibility of identifiable {\em dynamical} differences
between active and inactive hosts.

Removal of angular momentum and transportation of disk gas towards the
nucleus may be driven by external mechanisms, such as tidal
interaction or minor mergers, or internal instabilities such as bars
and lopsided disks.  HI is often the most spatially extended component
of a galaxy's disk so is sensitive to interactions and minor mergers;
because gas is dissipative, it also responds in a highly non-linear
way to small deviations from axial symmetry making it a valuable
tracer of barred or weak oval potentials \cite{ms99}.  HI synthesis
imaging provides unique information on the global mass distribution
and kinematics of the disk to large radii and a long-term dynamical
history of the galaxy; interactions and mergers alter disk gas masses,
bars drive gas inwards and even small disturbances leave {\em
kinematic} imprints. 

Ionised gas kinematics trace the gaseous response
closer to the nucleus, while stellar (ballistic) dynamics provide
independent constraints on the potential in the presence of strong
gaseous streaming. Here we describe 3D studies of the distribution and
kinematics of neutral and ionised gas and stars in active and inactive
galaxies.


\section{The {\bf V}LA  {\underline H}ydrogen  {\underline I}maging and  {\underline K}inematics of  {\underline IN}active  {\underline G}alaxies and  {\underline S}eyferts Survey (The {\bf VHIKINGS} Survey)}
 \label{subsec:sample}

\begin{figure}
\hspace{-6mm}    
\includegraphics[width=1.1\textwidth]{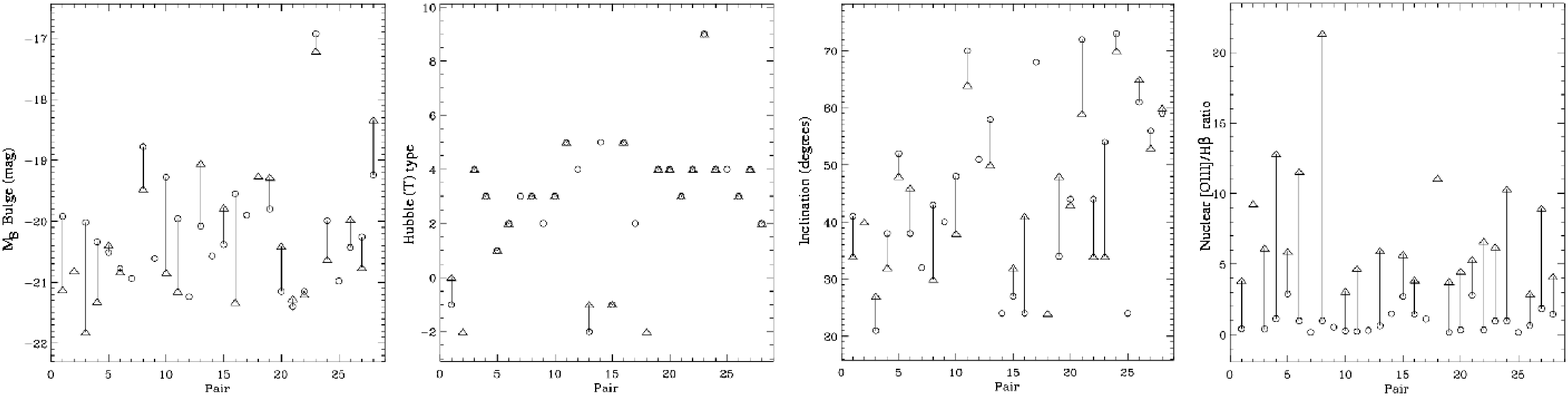}    
\caption{Bulge magnitude (M$_B$), Hubble type (T), inclination (i) and nuclear [O{\sc ii}]/H$\beta$ ratio for the VHIKINGS sample Seyfert ($\Delta$) and control ($\circ$) pairs.}
\label{sample}
\end{figure}

The VHIKINGS survey is a high angular resolution (20$''$) HI synthesis
imaging spectroscopic survey of 56 Seyferts and inactive controls
which aims to compare statistically the structural and dynamical
properties of active and inactive galaxies, specifically (a)
relationships between the presence of nuclear activity and disk
properties such as HI content, distribution, global and detailed
kinematics; (b) the gaseous environment of active and inactive
galaxies, such as the prevalence of gas-rich low optical brightness
dwarf galaxies.

Our master sample is selected from the RSA catalogue with complete nuclear
spectroscopic classification  \cite{hfs97} and comprises 39 Seyfert
galaxies, with V$_{sys}<4000~km~s^{-1}$, $B_T<12.5$ mag,
$20^{\circ}<i<70^{\circ}$, paired with 39 control galaxies with
carefully matched optical properties $B, V, i$ and Hubble type, T (RC3
classification plus visual confirmation of DSS images). For the
VHIKINGS survey, the brightest 26 Seyferts (absolute nuclear V-band
magnitudes $<$$-$17.2) and their corresponding 26 control galaxies
were selected from this master sample (Table 1), plus fainter
Seyferts, M51 and M81 and their controls. Figure \ref{sample}
shows a pair-wise comparison of some sample properties.

The VHIKINGS survey is motivated by single dish HI surveys of
Seyferts, which found $\sim$40\% of Seyferts  \cite{hbs78,mw84} showed
disturbed kinematic profiles different to those of normal galaxies and
a larger scatter in the distributions of M$_H$/L$_B$ and M$_H$/M$_D$
ratios vs. Hubble type, and detailed VLA HI studies of individual
Seyfert hosts, which revealed tidal tails and intra-group gas
 \cite{m95,mun01,mun03}, previously unknown dwarf galaxies close to,
but distinct from main disks \cite{m95} and non-linear gas dynamics,
such as streaming shocks in bars \cite{ms99}.

\begin{figure}
\includegraphics[width=\textwidth]{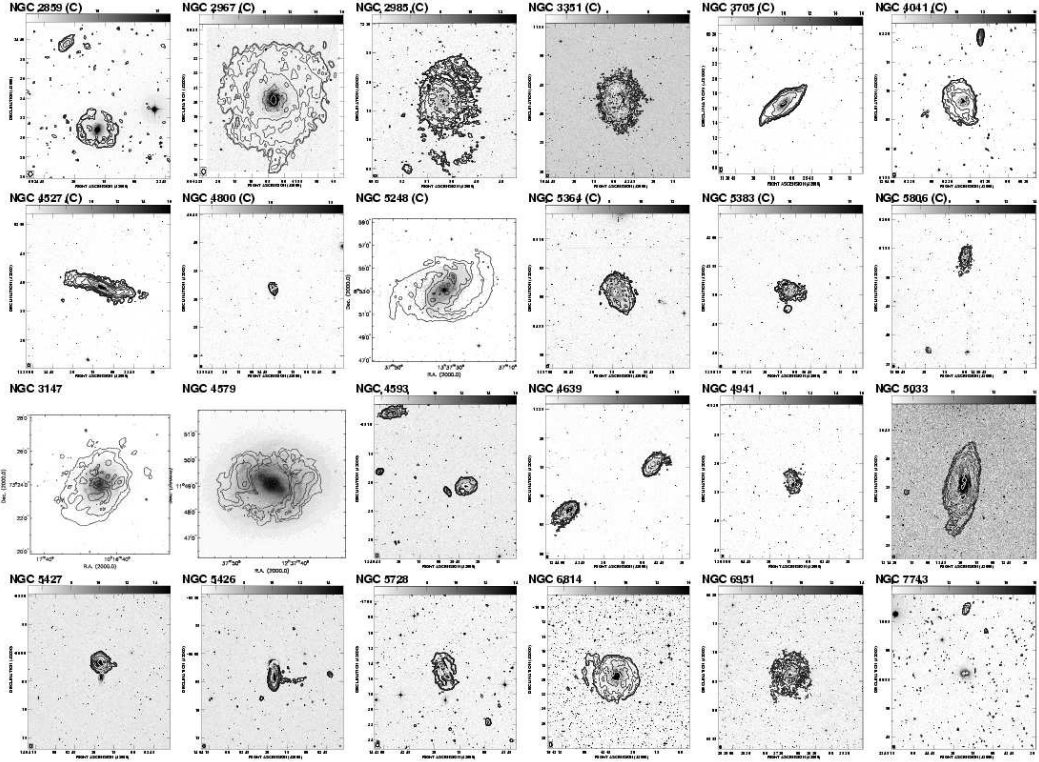}    
\caption{VLA HI images of a selection of galaxies from the VHIKINGS survey (HI intensity contoured on DSS R-band images).}
\label{HI}
\end{figure}

The VLA in C-configuration was used with an instrumental setup that
ensures sensitivity to column densities N$_{H}$$>$10$^{20}$~cm$^{-2}$
and a bandwidth corresponding to $\Delta$V=1100~km~s$^{-1}$; this
combination is proving valuable for accurate continuum subtraction and
the detection of gas-rich dwarf galaxies with low optical surface
brightness that are in the field of view and within the velocity range
to be associated with the target galaxies. These satellites are often
absent from archive data that fail to cover a sufficient velocity
range. Some of these satellites can be seen in Figure \ref{HI}, which
shows a selection of HI maps of Seyferts and controls (C) from the
survey. Wider comparisons with galaxies in the WHISP survey
 \cite{n05} is underway.

\section{Complementary Optical IFU Followup Studies}

\begin{figure} 
\includegraphics[width=\textwidth]{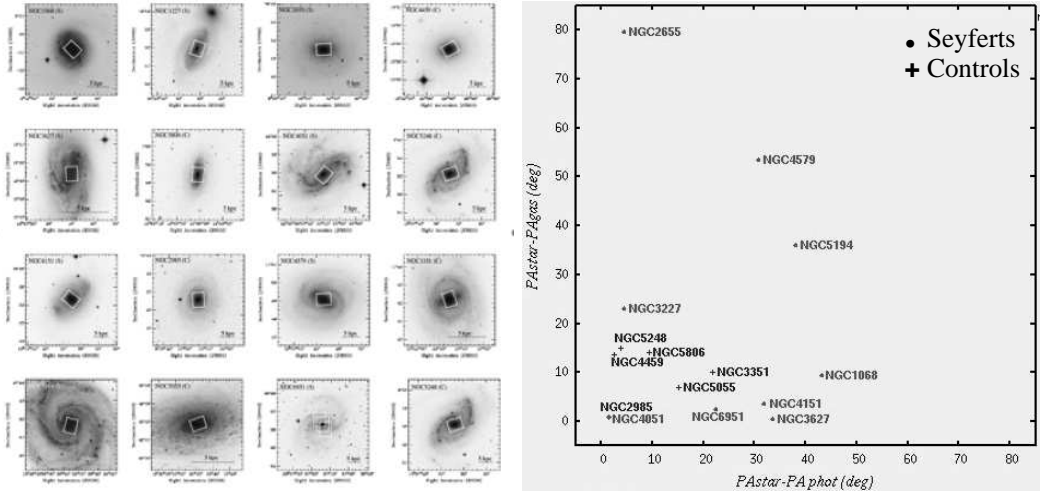} \caption{DSS
R-band images of SAURON subsample, where $\Box$ = observed regions;
Comparison of differences in mean position angle of photometric (phot)
and kinematic (stars and gas) axes, $PA(phot-stars)$ vs
$PA(stars-gas)$.}  \label{sauron} \end{figure}

We have used the SAURON IFU on the WHT to conduct a 3D imaging
spectroscopic study of stellar and gaseous distributions and
kinematics of the central kiloparsecs of a distance-limited
(V$_{sys}<1600~km~s^{-1}$) subsample of active and inactive galaxies
selected from the VHIKINGS sample (Dumas et al. in preparation). This
study was motivated by the clear inadequacy of single aperture spectra
or long-slit data or imaging alone in the presence of complex host
and nuclear kinematics.  The data form part of a multiwavelength
campaign to investigate host galaxy structure and kinematics on size
scales ever-closer to the nucleus, aimed at identifying or eliminating
possible AGN triggering and fuelling mechanisms (see also
Garc\'ia-Burillo these proceedings). This work builds on our previous
detailed studies of 2-D gaseous and stellar velocity fields of
individual Seyfert galaxies, such as NGC~4151 \cite{ms99,amp05},
NGC~2110 \cite{fer04} and NGC~1068 \cite{ems06} where evidence of
gaseous streaming and inflow in the inner few kpc was identified,
hinting at a mechanism for delivering gas to the circumnuclear
regions. Fig. \ref{sauron} shows DSS images of the SAURON subsample
(Left panel) and a comparison of photometric and kinematic axes
position angles. Comparisons are also being made with a wider range of
Hubble types from the SAURON survey of early type galaxies
 \cite{dez02,em04,fb06} and LINERs in elliptical hosts from Peletier
 \& Barthel (in prep.). Detailed modelling of the 2D stellar and
gaseous kinematics is underway to identify and quantify
deviations from circular motion that may be important for nuclear
fuelling, using both tilted ring modelling and Fourier decomposition
\cite[e.g.][]{sfdz97,wbb04,ems06} (see also Peletier these proceedings).

\end{document}